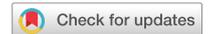

# Geometrical Constraints on the Tangling of Bacterial Flagellar Filaments

Maria Tătulea-Codrean & Eric Lauga

Many species of bacteria swim through viscous environments by rotating multiple helical flagella. The filaments gather behind the cell body and form a close helical bundle, which propels the cell forward during a "run". The filaments inside the bundle cannot be continuously actuated, nor can they easily unbundle, if they are tangled around one another. The fact that bacteria can passively form coherent bundles, i.e. bundles which do not contain tangled pairs of filaments, may appear surprising given that flagella are actuated by uncoordinated motors. In this article, we establish the theoretical conditions under which a pair of rigid helical filaments can form a tangled bundle, and we compare these constraints with experimental data collected from the literature. Our results suggest that bacterial flagella are too straight and too far apart to form tangled bundles based on their intrinsic, undeformed geometry alone. This makes the formation of coherent bundles more robust against the passive nature of the bundling process, where the position of individual filaments cannot be controlled.

There is arguably no complex geometrical structure more central to biology than the helix. Two helical strands of nucleic acid spiral around each other to form the blueprint of life[1], the arteries and vein that make up the human umbilical cord are twisted in a helical pattern[2], and our human hearts beat with a spiralling contraction of the myocardium[3]. Many large eukaryotic cells, particularly in plants, set up helical flows to enhance chemical transport within the cell[4] and some viruses also take this shape[5]. Prokaryotes too have learnt how to assemble blocks of proteins into a helical structure and put it to good use[6–9]. Indeed, many species of bacteria are able to propel themselves through the viscous medium they inhabit by exploiting an apparatus known as flagellum, whereby a relatively rigid helical filament is actuated by a specialised constant-torque rotary motor[10–12]. In the case of multi-flagellated bacteria, the filaments are swept behind the cell body and rotate together in a coherent bundle, which leads to an interval of straight swimming called a "run"[13]. To change swimming direction, at least one motor must switch its sense of rotation, upon which the associated flagellum will leave the bundle and generate an imbalance of forces. The subsequent reorientation of the cell is called a "tumble"[14].

This run-and-tumble behaviour lies at the heart of bacterial motility. It is known that the initial stage of bundling is enabled by an elastohydrodynamic instability of the hook[15], and from there onwards both the counter-rotation of the cell-body and hydrodynamic interactions between the filaments contribute to the synchronization and formation of the bundle[16–22]. Computational studies have investigated the effect of the number of flagella on the motility of the cell[23], and the role played by polymorphic transformations[24] and mismatched motor torques[25] in the bundling and unbundling process. Experiments using scale models have provided further insights into the bundling mechanism[26,27].

Evidently, the ability of a multi-flagellated bacterium to bundle and unbundle its flagella is crucial for its mobility[28]. However, the cell has no control over the exact position of individual filaments whilst bundling, so it is intriguing that a passive process can reliably lead to ordered bundles. One potential problem is the helical geometry of the filaments, which may tangle around each other, like the intertwined strands of DNA. Yet neither experiments, nor computational studies have reported cases of tangled bundles. Despite the helical structure of flagella and the stochastic fluctuations in the torque applied by motors, it appears that tangling is always prevented thanks to a combination of hydrodynamic interactions between the filaments, their elastic deformations, and the shape and arrangement of flagellar filaments over the cell body. In this paper we aim to isolate and better understand the effect of one of those agents – what can geometry on its own, without the mechanics, tell us about the issue of tangling?

Department of Applied Mathematics and Theoretical Physics, University of Cambridge, Cambridge, CB3 0WA, United Kingdom. e-mail: m.tatulea-codrean@damtp.cam.ac.uk; e.lauga@damtp.cam.ac.uk

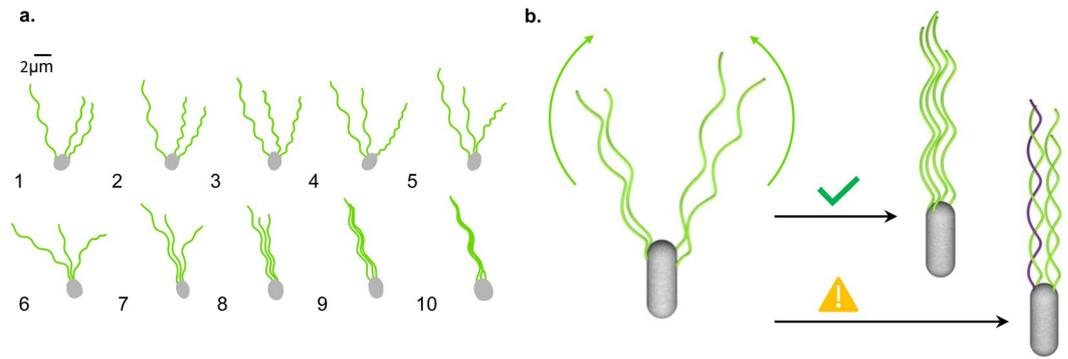

**Figure 1.** Bundling of flagellar filaments on a swimming bacterium. (**a**) Sketch of flagellar filaments rejoining a bundle from the proximal to the distal end of the filaments. Based on images of fluorescently labelled *Escherichia coli* in Turner, Ryu & Berg (2000) *J. Bacteriol.*, **182**, 2793–2801[13]. (**b**) Sketch of the problem, illustrating the possible outcomes of the bundling process. An ordered bundle can rotate continuously and propel the cell forward, whereas a tangled bundle cannot fulfill this purpose. One flagellar filament is drawn in a different colour to highlight that it revolves around the other filament – the two are said to be intertwined.

In this article, we show that the geometrical constraints imposed by the intrinsic helical shape of bacterial flagella are incompatible with the formation of tangled bundles, therefore increasing the robustness of the bundling process. The underlying assumption of this study is that the dynamics is quasi-steady, so that we may neglect elastic deformations and focus on the kinematics that can be achieved with the filaments in their intrinsic helical shape, as if they were perfectly rigid. Inspired by experimental observations, we consider the proximal to distal coming together of the filaments as they are being swept behind the cell body (see Fig. 1a) and we ask what are the possible bundles that can be obtained in this way (see Fig. 1b). We define a bundle to be the final configuration where all filaments are aligned with the cell body axis, and in particular we are interested in the possibility of forming a tangled bundle where one or more pairs of flagella are intertwined around each other, the precise mathematical definition being given later. It can be shown that a bundle of $N \geq 3$ parallel helices cannot be physically interlinked if no pair of helices is intertwined. Therefore, it is sufficient to consider the pairwise tangling of flagella within the bundle. Under the quasi-steady assumption, we establish the theoretical conditions under which a pair of rigid helical filaments may tangle, and then compare these constraints with experimental data collected from the literature. Our results suggest that bacterial flagella are too straight and too far apart to form tangled bundles based on their intrinsic, undeformed geometry alone.

## Theoretical Results

**Definition of tangled bundle.** In 1977, Robert M. Macnab published the first study discussing the geometry of bundles of helical filaments[29]. He examined two parallel identical helices of helical amplitude $r$, and found that they are intertwined (or tangled) if the distance between the helical axes is less than

$$d_c = 2r \sin(\Delta\chi/2), \tag{1}$$

or alternatively if the phase difference between the helices, taken without loss of generality to be in the interval $[0, \pi)$, is greater than

$$\Delta\chi_c = 2\sin^{-1}(d/2r). \tag{2}$$

In this case, as we scan the two filaments along their helical axes (from bottom to top in Fig. 1b), we observe that each filament revolves around the other in the same direction as the handedness of the helix (left-handed in Fig. 1b). Macnab then considered the same parallel helices, but anchored to a pair of fixed points (i.e. the location of the rotary motors). Given the rigidity of flagellar filaments, two tangled flagella could not be rotated continuously by the motors, as would happen during a run, due to an overwinding of the filaments around each other and an unsustainable build-up of torsion. Since a multi-flagellated bacterium could not swim with a bundle of tangled flagella, what restrictions are there on the shape and number of flagella in order to prevent tangling?

**Kinematic perspective of tangling.** We consider the tangling process in purely kinematic terms, i.e. as a series of configurations through which the filaments move in order to reach their final state, without considering the forces necessary to achieve this path dynamically. The only physical barrier guiding the tangling process is the condition that the two rigid structures cannot overlap at any snapshot in time. With this view in mind, we note that it is not possible to entangle two helical filaments by keeping them parallel and changing their phase difference, since they are physical objects that cannot cut through each other when the phase difference reaches $\Delta\chi_c$. Therefore, if flagella on a real bacterium were ever to form a tangled bundle, they would have to follow a sequence of non-parallel configurations. This could happen at the end of the tumble, when the flagella come back together.

In order to describe such behaviour, we begin by introducing the parameters that allow us to represent non-parallel helices. Since the hook is much shorter and much more flexible than the filament, we model the flagellum as being anchored to a fixed location by a universal point joint. We assume that the helices are tapered near their anchoring point, meaning that the helical amplitude $\tilde{r}(s) = (1 - e^{-\gamma s^2})r$ is a function of the height, $s$, along the axis[30]. This ensures that the filament is tangent to the helical axis at the anchoring point. In the body frame $\{\mathbf{e}_1, \mathbf{e}_2, \mathbf{e}_3\}$ of the helical filament, its centreline will be described by

$$\mathbf{r}_c(s) = \tilde{r}(s)\cos(2\pi s/p + \chi)\mathbf{e}_1 + \sigma\tilde{r}(s)\sin(2\pi s/p + \chi)\mathbf{e}_2 + s\mathbf{e}_3, \qquad (3)$$

where $p$ is the helical pitch, and $\sigma = \pm 1$ denotes the chirality. Note that the helix can be rotated by phase $\chi$ around the vector $\mathbf{e}_3$, which gives the direction of the helical axis. We use standard spherical polar coordinates and take the helical axis $\mathbf{e}_3$ parallel to the radial unit vector $\mathbf{e}_r(\theta, \phi) = (\cos\phi\sin\theta, \sin\phi\sin\theta, \cos\theta)$, whose components are given with respect to the lab frame $\{\mathbf{e}_x, \mathbf{e}_y, \mathbf{e}_z\}$. The two vectors spanning the plane orthogonal to the helical axis are $\mathbf{e}_1 = \mathbf{e}_\theta(\theta, \phi)$ and $\mathbf{e}_2 = \mathbf{e}_\phi(\phi)$. This setup is illustrated in Fig. 2. Therefore, the configuration of the helical filament is uniquely determined by the azimuthal angle $\phi$, the inclination angle $\theta$, and the phase $\chi$.

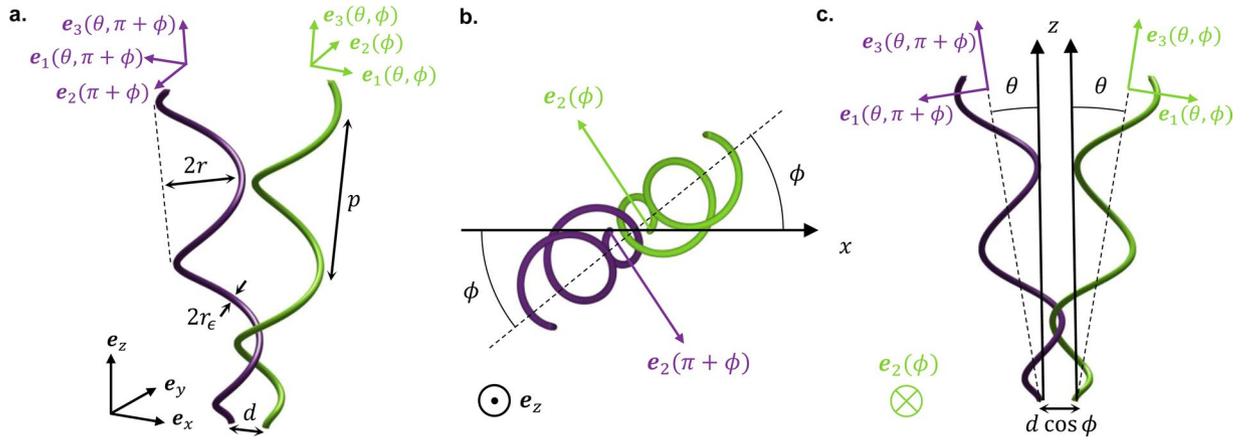

**Figure 2.** Geometry of tangling for two identical helices, each tapered near its anchoring point. (**a**) The easiest way to tangle two identical, rigid and helical filaments is to arrange them symmetrically about the midline between their anchoring points, which are a distance $d$ apart. In this case, one helix has configuration $(\phi, \theta, \chi)$, while the second has configuration $(\phi + \pi, \theta, \chi)$, obtained by rotating the first filament through an angle $\pi$ about the midline between their anchoring points. (**b**) Top view of (**a**), with $\mathbf{e}_z$ pointing out of the page. (**c**) Side view of (**a**), with $\mathbf{e}_2$ pointing into the page.

Now consider a pair of filaments anchored to two fixed points separated by a distance $d$. In the kinematic perspective, where we are only interested in the space occupied by the filaments at each moment in time, we can adaptively change our frame of reference $\{\mathbf{e}_x, \mathbf{e}_y, \mathbf{e}_z\}$ such that the anchoring points remain fixed and the direction $\mathbf{e}_x$ is always given by the line between the two anchoring points, despite the cell-body counter-rotations that may be observed in a stationary frame. Henceforth it is assumed that we are in a frame where the anchoring points, and hence the distance $d$, are fixed. Suppose we wanted to bring the two flagella from a configuration where they are not intertwined (e.g. pointing away from each other, as they would be at the end of a tumble) into a tangled bundle. Intuitively, the easiest way to achieve this is to arrange the helices symmetrically about the midline between their anchoring points at every step of the way, because this gives them space to circle around each other and intertwine, without blocking each other's way. This can be achieved by placing one helix in the configuration $(\phi, \theta, \chi)$ and the second in the configuration $(\phi + \pi, \theta, \chi)$, so that the helices are an image of one another under a rotational symmetry through angle $\pi$ about the midline between their anchoring points, as illustrated in Fig. 2. If tangling is not possible under these conditions, then it is not possible at all. In this symmetric configuration, the problem of intertwining two helices around each other becomes equivalent to intertwining one helix around a fixed vertical line of zero thickness, which represents the midline between the anchoring points of the two original helices. This is depicted in Fig. 3a. In our setup, the midline is the $z$-axis, and the helix is anchored to a point in the $xy$-plane at a distance $d/2$ away from the origin along the $x$-axis. Despite having zero thickness the vertical line has material properties, i.e. the filament cannot cross this line, because it would have to cross through its mirror image filament on the other side of the midline.

Since we want to determine the threshold where tangling is no longer possible, we must look at the limiting case where there is only enough space to tangle if the helical filament is touching the line of zero thickness, as seen in Fig. 3a. The trajectory of the filament as it tangles around the line can be characterised uniquely by a pair of coordinates $(s, \phi)$ indicating the contact point and the azimuthal angle over which the helical axis is inclined. In keeping with experimental observations illustrated in Fig. 1, we consider the problem as the contact point, $s$, slides up from the proximal to the distal end of the helix. As this happens, the portion of the helical filament below $s$ will gradually envelop the vertical line, until we reach a deadlock (see Supplementary Material online for videos of the tangling process). At a certain critical height $s_{\text{crit}}$, not only will the outside of the filament be in contact with the line at $s_{\text{crit}}$, by design, but the inside of the filament will also make contact with the line at height $s_{\text{crit}} - p$. Since the vertical line has material properties, any further increase of $s$ will lead to an overlap, so we cannot go beyond this point. This constraint is a consequence of the helical geometry of the filament and applies equally well to a helix of constant amplitude, so we start by deriving the threshold in the case of a perfect helix.

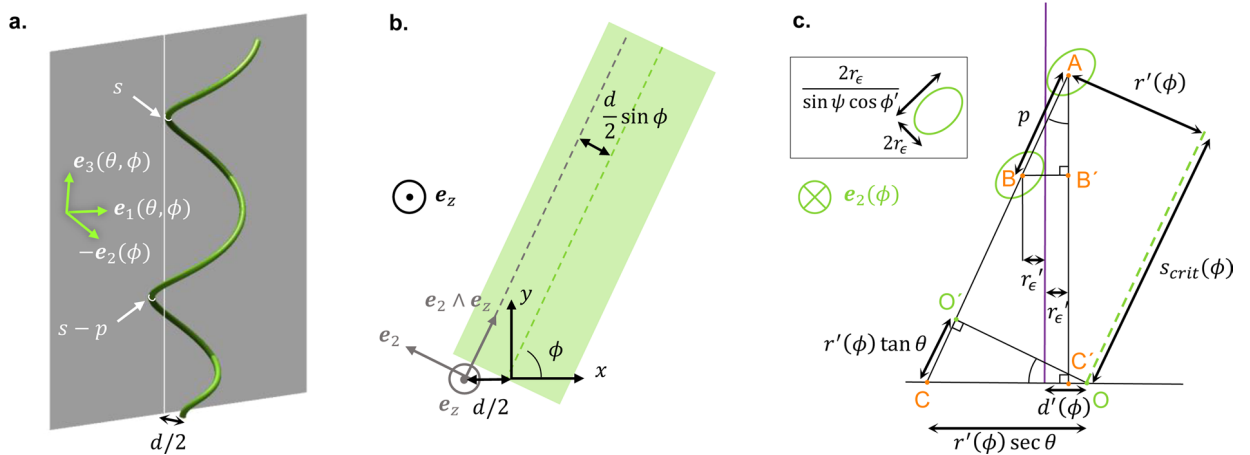

**Figure 3.** Computing the tangling threshold for two perfect helices. (**a**) In a symmetric configuration, the problem reduces to that of a rigid helical filament intertwining around a fixed vertical line of zero thickness (white). In the limiting case, tangling can be achieved by bringing the filament in contact with the line at point $s$ along the helix, and sliding this contact point up. Tangling cannot proceed past the point $s = s_{\text{crit}}$ where the exterior of the filament is in contact with the line at $s$, by design, but the filament makes contact with the line again on the interior at $s - p$. To locate this critical point we take a cross-section in the plane normal to $\mathbf{e}_2(\phi)$, indicated in this figure by a shaded screen, which we call the "focal plane". (**b**) Projection of (**a**) on the $xy$-plane. The projection of the cylindrical envelope containing the helix is shown in light green. The direction of the vector $\mathbf{e}_2 \wedge \mathbf{e}_z = (\cos \phi, \sin \phi, 0)$ is given by a dashed grey line. It is parallel to the projection on the $xy$-plane of the helical axis and lies at a distance $d \sin \phi / 2$ away from it. The focal plane is normal to the vector $\mathbf{e}_2(\phi) = (-\sin \phi, \cos \phi, 0)$ and goes through the origin, denoted by a grey dot. The helix is anchored to a fixed point at a distance $d/2$ away from the origin along the $x$-axis. (**c**) Sketch (not to scale) of the cross-section in the focal plane through a perfect helix of constant amplitude. The projection of the helical axis on the focal plane is drawn as a dashed light green line. The outline of the helical filament is also drawn in light green around the points A and B at heights $s_{\text{crit}}$ and $s_{\text{crit}} - p$ along the helical axis, respectively, and is tangent to the vertical line of zero thickness drawn in dark purple. The dimensions of the elliptical cross-section are shown in the top-left inset.

**Geometrical model for perfect helix.** We may calculate the critical point analytically, as a function of the azimuthal angle $\phi$ at which the helical axis is inclined, by taking a cross-section in the plane spanned by $\mathbf{e}_z$ and $\mathbf{e}_2(\phi) \wedge \mathbf{e}_z$, which we call the "focal plane" (see Fig. 3a,b). Recall that we are interested in finding the point of contact between the helix and the vertical $z$-axis. If one imagines keeping the azimuthal angle, $\phi$, fixed and varying the incline angle, $\theta$, in order to find the special value $\hat{\theta}(\phi, s)$ at which the helix is tangent to the $z$-axis at point $s$ along the helix, then the cylindrical envelope containing the helix will be "sliced" by the $z$-axis along the plane normal to $\mathbf{e}_2(\phi) = (-\sin \phi \cos \phi, 0)$ going through the origin, which explains our choice of focal plane. If we impose the further condition that the filament is tangent to the line at $s - p$ as well, we can determine the critical point $s_{\text{crit}}(\phi)$, and the critical incline angle $\theta_{\text{crit}}(\phi) = \hat{\theta}(\phi, s_{\text{crit}}(\phi))$.

The resulting geometric problem is sketched in Fig. 3c. The projected distance between the anchoring point and the $z$-axis is $d'(\phi) = d \cos \phi / 2$, and the projected distance between the filament centreline and the helical axis is $r'(\phi) = r \cos \phi'$, where $\phi' = \sin^{-1}(d \sin \phi / 2r)$ is a measure of the offset between the focal plane and the axis of the filament. The only quantity that requires careful consideration is the horizontal distance $r'_\varepsilon$ between the filament centreline and the $z$-axis. It can be shown that, for a small filament thickness $r_\varepsilon \ll r, p$ the intersection of the helical filament with the focal plane is well approximated by an ellipse with minor axis $2r_\varepsilon$ and major axis $2r_\varepsilon/\sin \psi \cos \phi'$, where $\psi = \tan^{-1}(2\pi r/p)$ is the helix angle. The major axis of the ellipse makes an angle $\theta + \tan^{-1}(\sin \phi' \tan \psi)$ with the $z$-axis. The tilting of the ellipse away from the vertical direction is partly due to the inclination angle $\theta$ applied to the entire helix, and partly due to the offset $\phi'$ between the focal plane and the helical axis, the latter effect being enhanced by the amount of sloping in the filament, as measured by $\psi$ (see Supplementary Material online for a full calculation of the cross-sectional shape). All these factors combined lead to a horizontal distance of approximately

$$r'_\varepsilon(\theta, \phi) \simeq r_\varepsilon \sqrt{1 + \frac{\sin^2(\theta + \tan^{-1}(\sin \phi' \tan \psi))}{\tan^2(\sin^{-1}(\cos \phi' \sin \psi))}} \tag{4}$$

between the filament centreline and the $z$-axis.

The multiple lengths in the problem can be linked together through the similarity of triangles $\triangle ABB'$ and $\triangle ACC'$, from which we derive

$$\sin \theta = \frac{2r'_\varepsilon(\theta, \phi)}{p} = \frac{r'(\phi)\sec \theta - d'(\phi) + r'_\varepsilon(\theta, \phi)}{r'(\phi)\tan \theta + s_{\text{crit}}(\phi)}. \tag{5}$$

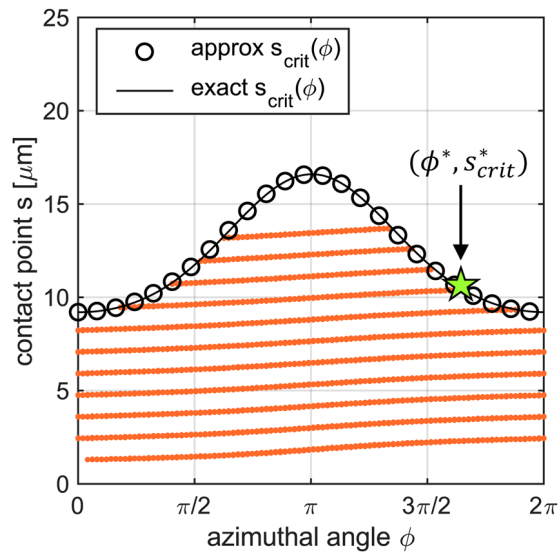

**Figure 4.** Representative example of the coordinate space ($\phi$, $s$) of configurations. The curve $s_{\text{crit}}(\phi)$ represents a constraint due to the helical shape of the filament. The approximate solution (circles) given by Eqs. (6) and (7) relies on a small $r_\varepsilon$ approximation. This is verified by the exact solution (solid black) where we solve the transcendental equations for $r'_\varepsilon(\theta, \phi)$ and $\theta_{\text{crit}}(\phi)$ iteratively. The sloping orange lines represent additional constraints due to the tapering of the helix and its anchoring to a fixed point, and they are computed numerically.

The first part of Eq. (5) will only be satisfied by a specific value of $\theta = \theta_{\text{crit}}$ dependent on $\phi$, for which we can solve in the limit of small thickness $r_\varepsilon \ll r$, $p$ using Eq. (4), to find that

$$\theta_{\text{crit}}(\phi) \simeq \frac{2r_\varepsilon}{p}\sqrt{1 + \frac{\sin^2(\tan^{-1}(\sin\phi'\tan\psi))}{\tan^2(\sin^{-1}(\cos\phi'\sin\psi))}} \tag{6}$$

Then, by rearranging the second part of Eq. (5), we obtain the critical height

$$s_{\text{crit}}(\phi) = \frac{p}{2} - \frac{d'(\phi)}{\sin\theta_{\text{crit}}} + \frac{r'(\phi)}{\tan\theta_{\text{crit}}} \tag{7}$$

This condition restricts trajectories in ($\phi$, $s$) coordinate space, as shown in Fig. 4. Because of this constraint, the tangling of a perfect helix around a fixed material line, or equivalently of two identical perfect helices around each other, is only possible for filaments shorter than $\max_\phi s_{\text{crit}}(\phi) = s_{\text{crit}}(\pi)$.

**Effect of tapering.** Beyond the case of helices with constant amplitude, tapering imposes further constraints on the ($\phi$, $s$) trajectory of the filament due to interactions near the anchoring point. The theoretical threshold for tangling from Eq. (7) is valid for a perfect helix with constant amplitude, $r$, meaning that the bottom end is not properly anchored to the point where the helical axis meets the $xy$-plane. Its locus is a sphere of radius $r$ around that point. If we wanted to intertwine the helix around the vertical line as much as possible, we could keep the azimuthal angle fixed at $\phi = \pi$, so that the helix is leaning towards the vertical line, and raise the contact point up to the global maximum of the constraint, which is $\max_\phi s_{\text{crit}}(\phi) = s_{\text{crit}}(\pi)$. As we do this, we must continuously decrease the phase $\chi$, for which we have the exact solution $\hat{\chi}(\phi, s) = \sin^{-1}(\sigma d \sin\phi/2r) - 2\pi s/p$, while also decreasing the incline angle $\hat{\theta}(\phi, s)$. The motion this would generate is that of a helix being screwed around a fixed vertical line, which is possible because the bottom end of the helix is free to loop around the vertical line. During this screwing motion, each time the contact point covers an interval of length $p$ above some arbitrary point $s_0 + p/2$, the point $s_0$ will loop around the vertical line once (see Supplementary Material online for further explanations and a video illustrating this process). This is because the incline angle $\hat{\theta}(\phi, s) < \hat{\theta}(\phi, s_0)$ if $s > s_0$, so the vertical line already lies inside the cylindrical envelope of the helix up to the point $s_0$. Meanwhile, the point $s_0$ revolves around this cylindrical envelope once every time the phase $\hat{\chi}(\phi, s)$ covers a period of $2\pi$, which happens each time the contact point $s$ covers an interval of length $p$.

Suppose we wanted to follow the same strategy (i.e. keep $\phi$ fixed and increase $s$) with a tapered helix that is anchored to a fixed point. Then every time the contact point goes up by one helical pitch, the filament would have to intersect the vertical line. This is because the bottom end of the anchored helix is fixed relative to the vertical line, but there exists some point $s_0$ higher up the filament which loops around the vertical line once. Since the filament is a continuous curve between these two points, it must intersect the vertical line in the process. These intersections can be determined numerically, with results shown as dotted orange lines in Fig. 4, and represent physical constraints to the motion of the filament.

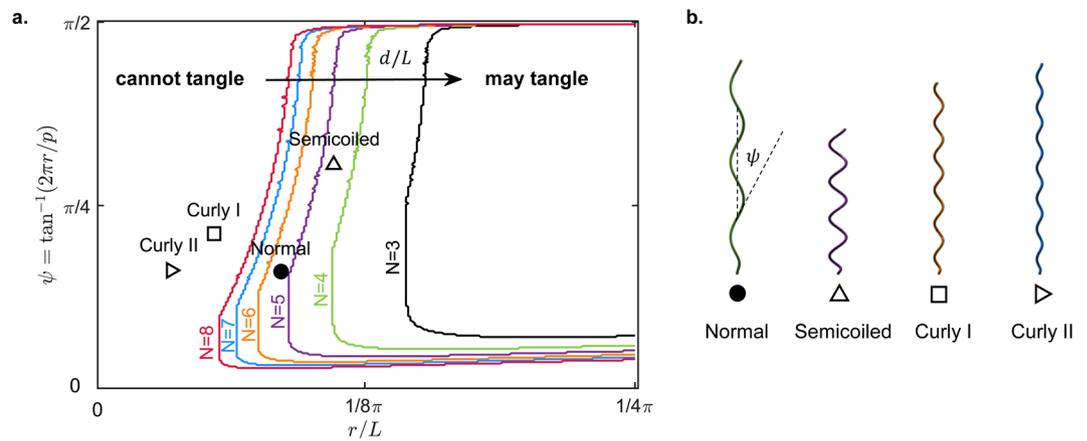

**Figure 5.** Single cell results. (**a**) Phase space for an identical pair of isometric filaments with constant length, $L$, and helical shape described by $(r/L, \psi)$. Each boundary is labelled with an integer number of flagella, $N$, and corresponds to a different partitioning of the phase space into two regions. The boundary for a given $N$ is obtained using a distance between the anchoring points equal to the expected minimum distance between any pair of flagella, when there are $N$ flagella distributed uniformly across the cell surface. To the left of each boundary, our theoretical model predicts that no pair amongst the $N$ flagella should be close enough to form a tangled bundle, while to the right of the boundary the closest pair of flagella may be able to form a tangled bundle. The values of the control parameters $(L, r_\varepsilon, l, w)$ are representative of the bacterial strain *E. coli* AW405[14]. (**b**) Isometric filaments in the shape of the four polymorphic forms most commonly seen in bacteria.

In order to complete tangling up to the highest possible point along the helix, we must navigate the coordinate space $(\phi, s)$ in such a way as to avoid hitting any physical constraints. We may work our way up between the sloping lines shown in Fig. 4 and around the $\phi$-periodic domain until we reach the constraint $s = s_{\text{crit}}(\phi)$ (see Supplementary Material online for further explanations and a video illustrating this process). The maximum height up to which two helices can tangle around one another is found numerically to be $s^*_{\text{crit}} = s_{\text{crit}}(\phi^*)$, where $\phi^*$ is a consequence of tapering. This threshold depends on all the parameters in our model, namely the geometry of the flagellum described by $(p, r, r_\varepsilon)$, and the distance, $d$, between the two anchoring points.

## Comparison with Experiments

What are the implications of these results for the morphology of real bacteria? For a single cell, with a given cell body size and a given number of flagella, the possibility of forming tangled bundles will be determined by the shape of those flagella. If one were to design a robotic micro-swimmer resembling multi-flagellated bacteria, one could consider a continuous parameter space for the helical shape of propellers, whereas in real bacteria the shape belongs to a discrete set of polymorphic forms. On the other hand, if the shape of the filament is assumed to be one of the polymorphic forms, the possibility of forming tangled bundles in a population of cells with stochastic variations in morphology will be determined by the relationship between cell body size and number of flagella, which affects how closely together the flagella are packed. We address both of these points in the next two subsections, using parameter values taken from the biological literature.

### Single cell results.
We start by considering a single cell of fixed dimensions, and we allow for isometric variations in the shape of the helical flagellum (i.e. continuously varying pitch and radius, at constant integrated length). Any helical filament of fixed length $L$ can be described by a pair of coordinates $r/L$, the helical radius relative to the length of the filament, and $\psi = \tan^{-1}(2\pi r/p)$, the helix angle. The coordinate space $(r/L, \psi)$ can be divided into two regions where the pair of filaments may or may not tangle, according to whether the theoretical threshold, $s^*_{\text{crit}}(p, r, r_\varepsilon, d)$, is greater or less than the full height of the filament, $s_{\text{end}}(p, r, L)$. The boundary between these two regions depends on $d/L$, the distance between the anchoring points relative to the length of the filaments, which is our control parameter.

In order to choose realistic values for $d/L$, we note that the expected minimum distance between any two flagella on a bacterium will depend on the size of the cell body and the number of flagella. Modelling the cell body as a capsule (i.e. a cylinder with two hemispherical caps) of length, $l$, and width, $w$, we compute the expected minimum distance between any two rotary motors, $d_{\min}(N, l, w)$, by simulating 10,000 random arrangements of $N$ motors on the cell body sampled from a uniform distribution over a capsule (this sample size is sufficient to make our estimate for $d_{\min}$ accurate to within 1%). The expected minimum distance is then scaled by the mean length of a flagellum reported in the literature. Our theoretical predictions for a single cell, using the average cell body size and average length of flagella for the bacterial strain *E. coli* AW405[13,14], are shown in Fig. 5a. As the number of flagella increases and the relative separation between flagella decreases, the region where filaments could, in theory, form a tangled bundle occupies increasingly more of the parameter space, as expected.

The four most common polymorphic forms, depicted in Fig. 5b, are indicated as discrete points in the continuous parameter space of Fig. 5a. We observe that the normal form, the most common of the polymorphic forms, does not enter the region where tangled bundles are possible until the number of flagella is greater than or equal to six. This compares favourably with the average number of flagella for *E. coli* AW405 quoted in the literature, which is between three and four flagella[13]. From the phase space we also deduce that the semicoiled form is the most susceptible to tangling. Finally we note that, at a typical number of three or four flagella[13], all four polymorphic forms lie in the region where the intrinsic geometry of the flagellar filaments should not allow the formation of tangled bundles.

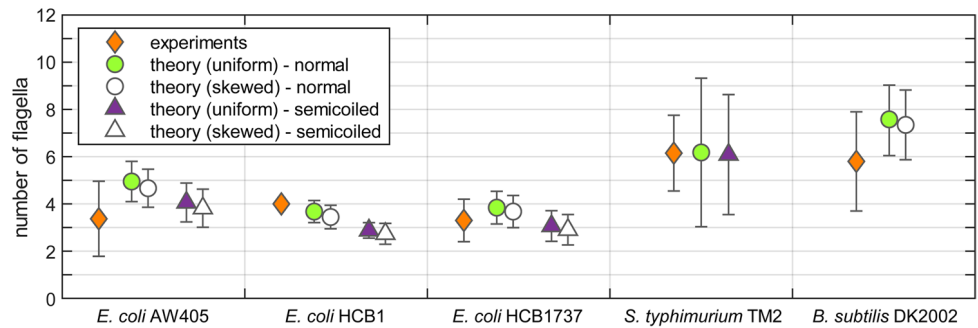

**Figure 6.** Comparison between theoretical predictions and experimental data for three multi-flagellated species of bacteria: *Escherichia coli*, *Salmonella typhimurium* and *Bacillus subtilis*. Our predicted critical number of normal flagella (circles) and semicoiled flagella (triangles) is compared against the real number of flagella counted in experiments (diamonds). For the filled circles and triangles we have assumed that flagella are uniformly distributed over the cell body, while for the empty symbols we have used a non-uniform distribution that is skewed towards one pole of the cell (for *Escherichia coli* and *Bacillus subtilis*, this data was available in the literature[31,44]; no equivalent data was available for *Salmonella typhimurium*). The error bars represent one standard deviation.

**Population level results.** We now consider stochastic variations in the shape of the cell body and the length of flagella, but keep the helical shape of the filaments constant. We model the cell body as a capsule, as before, and we generate a sample of 500 cells with independent normally distributed body lengths, $l$, and widths, $w$, with means and variances taken from the literature. For each of these cells, we compute the expected minimum distance between any two rotary motors, $d_{\min}(N, l, w)$, by randomly placing $N$ motors on the cell body, distributed either uniformly or according to a known asymmetric distribution[31,44], and averaging over 10,000 realisations. The flagellar filaments attached to these motors have a given geometry ($p, r, r_\varepsilon$) corresponding to one of the bacterial polymorphic forms found in nature, but their length is a random variable, sampled either from an experimentally measured distribution[32] or from a log-normal distribution if only the mean and variance are available in the literature. For each simulated cell we compute the theoretical threshold, $s^*_{\text{crit}}(p, r, r_\varepsilon, d_{\min}(N, l, w))$, and compare this with the length of the simulated filaments to determine if, for that number of motors, we would expect any two flagella to be sufficiently near each other to form a tangled bundle. This allows us to predict the maximum number of flagella that the cell may have without running the risk of tangling.

Our theoretical predictions, based on parameter values taken from the literature[13,14,32–43], are shown in Fig. 6 (circles and triangles) alongside the real number of flagella counted in experiments (diamonds), for the three most widely studied species of multi-flagellated bacteria: *Escherichia coli*, *Salmonella typhimurium* and *Bacillus subtilis* (see Supplementary Material online for data sources). For our theory, we use the two most common polymorphic forms that the bacterial flagellum can take, namely the left-handed normal form used in runs (circles) and the right-handed semicoiled form used in tumbles (triangles)[41]. The empty symbols represent an estimate based on a non-uniform arrangement of molecular motors over the cell body, according to distributions reported in the literature for *Escherichia coli* and *Bacillus subtilis*[31,44] (no equivalent data was available for *Salmonella typhimurium*). These estimates are always slightly lower than those for a uniform distribution, because the effective distance between flagella is reduced when the molecular motors are crowded towards one pole of the cell, which makes tangling more likely. However, the relative difference between the two estimates is very small because the asymmetry in motor distribution is not very pronounced.

Remarkably, we find that the theoretical thresholds in Fig. 6 are always above, or less than one below, the real number of flagella observed in experiments. This close agreement strongly suggests that the morphologies of the flagellar filaments and of the cell body in the strains we have considered make it unlikely for flagella to form tangled bundles under a quasi-steady motion without elastic deformations. We note that two other polymorphic forms (curly I and II) are sometimes observed in the flagellar bundles of swimming bacteria[13], but they are so slender that the corresponding theoretical thresholds are much higher than for normal and semicoiled flagella, as evidenced in Fig. 5. Hence, preventing their tangling does not appear to impose strong physical constraints on the cell.

**Discussion**

The number of flagella is a fundamental control parameter for multi-flagellated bacteria, and its connection to bacterial spreading and speed of locomotion has been previously investigated[23,45]. In the current study, our kinematic model for tangling allows us to examine whether the number of flagella could be linked to the robustness of bundling. By comparing the thresholds from our theoretical model with the real number of flagella counted in experiments, we deduce that the intrinsic geometry of flagellar filaments should not allow the formation of tangled bundles. This is consistent across the four most common polymorphic forms of the bacterial flagellum, and across three different strains of *Escherichia coli* and one strain of *Salmonella typhimurium* for which we could test a consistent set of data. We note that in the case of *Bacillus subtilis*, the geometrical constraints identified in this paper and presented in the last column of Fig. 6 are only suitable for the shorter mutant DK2002[38], which is equipped with fewer flagella. In contrast, the wild-type and highly-flagellated strains of *Bacillus subtilis* can have up to forty flagella with only a slight increase in cell body size (less than 50%), meaning that the constraints imposed by the intrinsic geometry of flagella are not sufficient to rule out tangling. The thresholds predicted by our model for the wild-type strain DS9540[38] were of approximately nine normal flagella or thirty curly flagella, compared to experimental measurements of twenty-six derived from a basal body count[44].

For *Escherichia coli* and *Salmonella typhimurium*, we can formulate our conclusion either in terms of the helical geometry of the polymorphic forms, which appears to be too slender to allow tangling (Fig. 5), or in terms of the relative spacing between motors, which appears to be too wide to allow tangling (Fig. 6). Alternatively, we can think about the space of configurations as being partitioned into different regions by the physical constraint that the filaments cannot overlap (Fig. 4). Using parameter values from the experimental literature, our theoretical model suggests that tangled bundles are isolated in configuration space from the state in which flagella typically find themselves at the end of a tumble (see Fig. 1b). Since tangled bundles are physically inaccessible, this means that the bundling process is more robust against the lack of coordination between molecular motors and the inability of the bacterium to control the individual trajectory of each filament.

There are two geometrical aspects which we have not included in the present study. Firstly, the setup described in Fig. 2 consists of filaments anchored to a pair of fixed points in free space. Although we model the geometry of the cell body in order to determine the typical spacing between filaments, the surface of the cell body is not included as a physical constraint in our configuration space from Fig. 4. However, we note that the typical distance between flagella before their intrinsic geometry allows them to tangle in free space is between 0.1 $\mu$m for curly and 0.5 $\mu$m for semicoiled, whereas at the poles of the cell the radius of curvature is around 0.35–0.6 $\mu$m depending on the strain. Hence, the effect of surface curvature will be most important for semicoiled flagella attached to smaller cells, since the surface could then be bulging out between the anchoring points at an angle as high as 46° from the horizontal, while the helical filament of semicoiled makes an angle of approximately 35° with the horizontal. In this case, the steric interactions between the filaments and the cell body surface will depend greatly on the details of the tapering of the helix near the anchoring point. For normal flagella, the cell body surface would be raised at an angle of at most 34° to the horizontal near the anchoring point, even on the smallest cells we have considered, whereas the filament of normal flagella makes an angle of 61° to the horizontal. Hence, the presence of the curved surface will have very little effect on the tangling of normal flagella, and even less so for curly I and II.

The second aspect of the geometry which we have neglected is that the shape of the filament can change dynamically while the cell tumbles due to polymorphic transformations which propagate from the proximal to the distal end of the filament[13,24,41]. The angle between sections of the filament in different polymorphic states is given by Hotani's rule[46]. This aspect of the geometry requires a fully dynamical model including the elasticity of the filament and the hydrodynamic load that triggers the transformation, and goes beyond the scope and purpose of the current study.

Furthermore, our modelling approach is purely kinematic and neglects a number of important dynamical factors. On one hand, elastic deformations might allow the entanglement of longer filaments than we have predicted but, on the other hand, thermal fluctuations would increase the effective thickness of the filaments, thereby decreasing the likelihood of tangling. The finite length of the hook is another element of complexity that cannot be addressed in our simple model, because it would introduce further degrees of freedom. The end of the filament would no longer have to be anchored to a fixed point, but could hover within a small distance (the length of the hook) away from the true location of the molecular motor, which is fixed. When the filaments are being swept behind the cell body, the effective distance between the proximal ends of the filaments could be reduced, making it easier to form a tangled bundle.

Hydrodynamic interactions play a very important role in the formation of stable bundles since they favour the synchronization of flagella, leading to helical filaments rotating in phase with one another. The clockwise rotation of normal flagella, together with the counter-rotation of the cell body, lead to a right-handed wrapping of the filaments inside the bundle, as seen in experiments[26] and computations[19,21]. This kind of wrapping is made possible by the elastic deformation of the filaments and is different to the left-handed tangling considered in this article, which is the only one allowed by the left-handed intrinsic geometry of the normal flagellum[29]. Simulations have shown that, if hydrodynamic interactions are switched off, the bundle comes apart because the filaments are not geometrically interconnected due to the right-handed wrapping[19]. In this article we have focused instead on the tangling which arises from the intrinsic geometry of the filaments (left-handed for normal flagella, and right-handed for the others) because it is the one that hinders the continuous actuation of the filaments during a run and the dispersal of the filaments during a tumble.

Our kinematic model for tangling is based on the idea that the set of hydrodynamically feasible paths is a subset of all possible geometries of motion. Thus, we have shown that hydrodynamics is not essential to prevent tangling in the quasi-steady regime – geometry imposes sufficient restrictions to avoid the formation of tangled bundles. If we relax the quasi-steady approximation, the set of possible geometries of motion will expand greatly since the shape of the filament could undergo arbitrary elastic deformations. Then tangling will (obviously) always be theoretically possible within the set of all possible geometries of motion. It will therefore be necessary to identify the filament kinematics that is also dynamically feasible, given the background flow conditions and the input of the rotary motors. Such a wide range of dynamic conditions is computationally inexhaustible, which is where our model can provide a stepping stone for further studies and narrow down the search. Based on the intrinsic geometry of the filaments, we have identified the kinematics that leads to tangling (see Supplementary Video S3). We compared this kinematics to the typical frequency of rotation of bacterial motors, and we identified that the motors would have to switch the direction of rotation one hundred times per second and spend an equal amount of time rotating in either direction, whereas in real bacteria the motors only switch the direction of rotation once per second and spend the vast majority of the time rotating clockwise[14]. The qualitative difference between the kinematics that leads to tangling and that of real flagella is a further argument for the robustness of the bundling process.

Our results bring a new perspective on the ability of multi-flagellated bacteria to passively form coherent, tangle-free bundles of flagella that can be continuously actuated during a run and easily taken apart during a tumble. At one extreme, we know that straight flagella could not intertwine, but they would also not be able to propel the cell forward in a viscous fluid. It appears that the polymorphic forms most commonly adopted by bacterial flagella are sufficiently coiled to be efficient propellers[47] but also sufficiently slender to avoid tangling.

### Code availability
The custom codes used in this study are available from the corresponding authors upon request.

### Acknowledgements

We thank the anonymous reviewers for their useful comments and suggestions. We also thank Alexander Chamolly, Debasish Das and Anne Herrmann for valuable comments on an early version of the manuscript. We gratefully acknowledge funding from the George and Lillian Schiff Foundation (award to M.T.C.) and the European Research Council under the European Union's Horizon 2020 research and innovation programme (grant agreement 682754 to E.L.).

### Author contributions

M.T.C. and E.L. designed the research, M.T.C. performed the analytical calculations and numerical simulations. M.T.C. and E.L. analysed and interpreted the data and wrote the manuscript.

### Competing interests

The authors declare no competing interests.

### Additional information

**Supplementary information** is available for this paper at https://doi.org/10.1038/s41598-020-64974-6.

**Correspondence** and requests for materials should be addressed to M.T.-C. or E.L.

**Reprints and permissions information** is available at www.nature.com/reprints.

**Publisher's note** Springer Nature remains neutral with regard to jurisdictional claims in published maps and institutional affiliations.